\documentclass[a4paper,14pt]{article}
\usepackage[english]{babel}
\pdfoutput=1 

\usepackage{jheppub} 

\usepackage[T1]{fontenc} 
\usepackage{comment}

\usepackage[bbgreekl]{mathbbol} 

\usepackage{bbold}
\usepackage{mathbbol}
\usepackage{bbm}

\usepackage{pdfrender}

\DeclareSymbolFont{eulerletters}{U}{eur}{m}{n}
\DeclareMathSymbol{\callambda}{\mathord}{eulerletters}{"15}

\makeatletter \@addtoreset{equation}{section} \makeatother

\def\slashchar#1{\setbox0=\hbox{$#1$}           
	\dimen0=\wd0                                 
	\setbox1=\hbox{/} \dimen1=\wd1               
	\ifdim\dimen0>\dimen1                        
	\rlap{\hbox to \dimen0{\hfil/\hfil}}      
	#1                                        
	\else                                        
	\rlap{\hbox to \dimen1{\hfil$#1$\hfil}}   
	/                                         
	\fi}

\pdfoutput=1

\title{\boldmath Harmonic superspace and Kaluza–Klein reduction}

\author[]{Nikita~Zaigraev}

\affiliation[]{Bogoliubov Laboratory of Theoretical Physics, JINR,\\
	Joliot-Curie st. 6, 141980 Dubna, Moscow region, Russia}
\affiliation[]{Moscow Institute of Physics and Technology,\\ Institutskiy per. 9, 141700 Dolgoprudny, Moscow region, Russia}

\emailAdd{nikita.zaigraev@phystech.edu}

\abstract{
We extend the Kaluza--Klein reduction procedure to the framework of harmonic superspace and apply it to the construction of off-shell superfield formulations of supersymmetric gauge multiplets. The reduction is implemented by a deformation of the harmonic derivatives by mass terms, while ensuring that manifest off-shell supersymmetry and Grassmann analyticity are preserved at all stages. As a first application, we reduce the free \(5D,\mathcal{N}=1\) Maxwell theory and obtain an off-shell superfield description of the BPS \(4D,\mathcal{N}=2\) massive vector supermultiplet. The resulting formulation naturally incorporates an analytic Stückelberg-type gauge symmetry. We then reduce linearized \(5D,\mathcal{N}=1\) Poincar\'e supergravity and derive the unconstrained harmonic-superspace action for the BPS \(4D,\mathcal{N}=2\) massive supergravity multiplet.
The technical simplicity of the procedure stems from the structure of gauge theories in harmonic superspace, where the gauge symmetry is naturally encoded in analytic prepotentials and harmonic derivatives. 
 The approach presented in this paper opens a direct
route to gauge-invariant off-shell superfield
formulations of massive higher-spin
$4D,\mathcal{N}=2$ BPS multiplets.  Furthermore, the
resulting theories can be reformulated in
$4D,\mathcal{N}=1$ superspace, opening the way
to gauge-invariant off-shell descriptions of $4D,\mathcal{N}=1$ massive higher-spin theories.

\medskip

\textit{\begin{flushright}
		Dedicated to Evgeny Ivanov on the occasion of his 80th birthday
\end{flushright}}
}

\makeatletter
\gdef\@fpheader{}
\makeatother

\begin{document} 
\maketitle
\flushbottom

\section{Introduction}

Harmonic superspace \cite{Galperin:1984mln, Galperin:1984av, 18} is a natural framework for constructing off-shell formulations of supersymmetric theories with eight supercharges.
Within the harmonic approach, off-shell formulations have
been developed for the hypermultiplet \cite{Galperin:1985de}, super Yang–Mills theory \cite{Galperin:1985bj,Galperin:1985va, Zupnik:1987vm}, supergravity \cite{Galperin:1987em, Galperin:1987ek, Bagger:1987rc,  Sokatchev:1988aa,  Zupnik:1998td, Butter:2015nza, Ivanov:2022vwc, Zaigraev:2026gjb, Ivanov:2024gjo}, and higher-spin theories \cite{Buchbinder:2021ite, Buchbinder:2025yef, Buchbinder:2025ceg, Buchbinder:2024xll}, including their interactions \cite{Buchbinder:2022kzl, Buchbinder:2022vra, Buchbinder:2024pjm, Zaigraev:2024ryg,  Zaigraev:2025xqo, Zaigraev:2026jvu,Zaigraev:2026vxf, Zaigraev:2026xzr}.
It is important that the formulation of all these theories is based on the fundamental principle of \textit{Grassmann analyticity}.

Despite this progress, the off-shell superfield description
of massive supermultiplets within harmonic superspace
remains relatively unexplored. The known examples include
the massive hypermultiplet
\cite{18, Ohta:1985ba}, which necessarily belongs to a
short (BPS) representation with $|m|=|Z|$, and the long
massive $4D,\mathcal{N}=2$ vector multiplet
\cite{Maslikov:1991yk, Khelashvili:1991er,
	Maslikov:1992rn, Galperin:1984av}.
In particular, gauge-invariant off-shell superfield
formulations of massive $\mathcal{N}=2$ supergravity and
higher-spin multiplets have not yet been established.

\smallskip

Massive higher-spin theories have been studied extensively.
The first massive higher-spin models were constructed by
Singh and Hagen \cite{Singh:1974qz, Singh:1974rc}.
Subsequently, Zinoviev developed gauge-invariant
formulations involving additional Stueckelberg fields
\cite{Zinovev:1983kh, Zinoviev:2001dt, Zinoviev:2002xn,
	Klishevich:1997pd}; see also
\cite{Buchbinder:2005ua, Metsaev:2003cu, Metsaev:2006zy}.
Related constructions include massive supergravity and
higher-spin supermultiplets
\cite{Zinoviev:2002xn, Zinoviev:2007js}.
Gravitational interactions of massive fields have also
been investigated
\cite{Zinoviev:2025jff, Zinoviev:2025tpx,
	Zinoviev:2025yqa, Zinoviev:2026azg}, as well as the
interaction of a massive spin-2 supermultiplet with
supergravity \cite{Zinoviev:2026ahi}.
Other approaches to massive higher-spin theories include
the frame-like formulation
\cite{Zinoviev:2008ze, Khabarov:2019dvi,
	Ponomarev:2010st, Zinoviev:2026osh}, the BRST approach
\cite{Pashnev:1997rm, Buchbinder:2025bzm,
	Buchbinder:2008ss}, and Kaluza--Klein dimensional
reduction \cite{Pashnev:1989gm, Aragone:1987dtt,
	Lindwasser:2023zwo, Fegebank:2024yft}.

\smallskip

At the superfield level, various $4D,\mathcal{N}=1$
massive theories have also been constructed, including linearized
massive supergravity
\cite{Gregoire:2004ic, Buchbinder:2002gh, Gates:2006cq},
massive supergravitino \cite{Buchbinder:2002tt}, and
massive arbitrary half-integer superspin multiplets
\cite{Koutrolikos:2020tel}.
Gauge-invariant superfield formulations of massive
$4D,\mathcal{N}=1$ higher-spin theories, however, remain
an open problem.
More recently, gauge-invariant formulations of massive
$3D,\mathcal{N}=2$ higher-spin supermultiplets,
corresponding to short (BPS) representations, were
constructed by means of Kaluza--Klein reduction from
massless $4D,\mathcal{N}=1$ higher-spin theories~\cite{Buchbinder:2026oou, Kuzenko:1993jp,
	Kuzenko:1993jq}.

\smallskip

These developments motivate the search for a systematic
superfield approach to massive higher-spin theories in
four dimensions, with both $\mathcal{N}=1$ and
$\mathcal{N}=2$ supersymmetry. A natural strategy is to
construct such theories by dimensional reduction from
five dimensions. Since $5D,\mathcal{N}=1$ supersymmetry
has eight real supercharges, harmonic superspace provides
a suitable framework for this construction
(see, e.g., \cite{Zupnik:1999iy}).
Unlike the component reduction approach, this framework preserves both manifest supersymmetry and \textit{Grassmann analyticity} —the fundamental geometric principle underlying all theories with extended supersymmetry—at every stage of the derivation.

\smallskip

In this paper, we study  Kaluza--Klein
reduction procedure in harmonic superspace and apply it
to the construction of massive $4D,\mathcal{N}=2$
supermultiplets. As we will see, this method is based on a deformation
of the $4D$ harmonic derivatives by the masses associated
with the Fourier modes. We first apply it to the free
$5D,\mathcal{N}=1$ Maxwell theory and obtain an
off-shell superfield description of the massive BPS
vector multiplet. We then extend the construction to
linearized $5D,\mathcal{N}=1$ Poincare supergravity,
deriving the corresponding off-shell action for the
massive BPS supergravity multiplet.

\smallskip

The paper is organized as follows. In section  \ref{sec: 2} we introduce our superspace notations and discuss reduction procedure for super Poincar\'e algebra. 
Section \ref{sec: 4} is devoted to dimensional reduction of hypermultiplet. 
In section \ref{sec: long} we discuss structure of long $\mathcal{N}=2$ vector supermultiplet. 
In section \ref{sec: 5} we present detail analysis of dimensional reduction of free $5D, \mathcal{N}=1$ vector multiplet theory. In section \ref{sec: 6} we illustrate how reduction work for linearized Poincar\'e supergravity and derive gauge-invariant superfield action for short (or BPS) $4D, \mathcal{N}=2$ massive supergravity multiplet.

\section{Superspace setup and dimensional reduction}\label{sec: 2}

Following Ref. \cite{18}, we begin by reviewing the compactification procedure
for the Poincar\'e superalgebra.
Since all $5D$ and $4D$ Poincar\'e superalgebras with 8 supercharges can be obtained from $6D, \mathcal{N}=(1,0)$, we begin our consideration with this general case.
$6D, \mathcal{N}=(1,0)$ Poincar\'e superalgebra without central charge is given by
\begin{equation}
	\{
	Q^i_{\hat{\alpha}}, Q_{\hat{\beta}j}   
	\}
	=
	2 \delta^i_j \gamma^{\hat{a}}_{\hat{\alpha}\hat{\beta}} P_{\hat{a}},
	\qquad
	P_{\hat{a}} = - i\partial_{\hat{a}}.
\end{equation}	
This algebra can be rewritten with manifest $5D$ Lorentz symmetry. Then we obtain $5D$ Poincar\'e superalgebra with real central charge $Z = -\partial_6$
\begin{equation}
	\{
	Q^i_{\hat{\alpha}}, Q_{\hat{\beta}j}   
	\}
		=
	2 \delta^i_j \gamma^{\hat{a}}_{\hat{\alpha}\hat{\beta}} P_{\hat{a}}
	+
	2  \delta^i_j \epsilon_{\hat{\alpha}\hat{\beta}} Z
\end{equation}	
One can also rewrite this algebra in a way in which only $4D$ Lorentz symmetry is manifest. We replace spinor indices as $\hat{\alpha} = (\alpha, \dot{\alpha})$, supercharges as
$Q_{\hat{\alpha}}^i = (Q_{\alpha}^i, \bar{Q}_{\dot{\alpha}}^i)$
and obtain $4D, \mathcal{N}=2$ superalgebra with complex central charge $Z = -i (\partial_5 - i \partial_6)$\footnote{Here following standard conventions we replace $\partial_4 \to \partial_6$.}:
\begin{equation}
	\begin{split}
	&\{Q^i_\alpha, \bar{Q}_{\dot{\alpha} j} \}
	=
	2 \delta^i_j \sigma^a_{\alpha\dot{\alpha}} P_a,
	\\
	&
	\{ Q^i_\alpha, Q^j_{\beta} \} = 2 \epsilon^{ij} \epsilon_{\alpha\beta} Z,
	\\
	&
	\{\bar{Q}_{\dot{\alpha} i}, \bar{Q}_{\dot{\beta}j} \} = - 2 \epsilon_{ij} \epsilon_{\dot{\alpha}\dot{\beta}} \bar{Z}.
	\end{split}
\end{equation}	
We see that central charges appearing in $5D$ and $4D$ superalgebras can be interpreted as extra components of the momentum in a higher-dimensional space.

Let us suppose that we have some (super)field  representation of $6D, \mathcal{N}=(1,0)$ super Poincar\'e algebra with fixed mass value
\begin{equation}
	P^{\hat{a}} P_{\hat{a}} = M^2.
\end{equation}	
 Then, if we reduce this theory to $5D$ ($\mathbb{R}^{6|8} \to \mathbb{R}^{5|8}\times S^1$) or $4D$ ($\mathbb{R}^{6|8} \to\mathbb{R}^{4|8}\times S^1 \times S^1$), we obtain the following relations for mass values
 \begin{equation}
 	M^2_{5D} = M^2 + Z^2, 
 	\qquad
 	M^2_{4D}  = M^2 + |Z|^2 
 \end{equation}	
  So if one considers 6D/5D massless theory, and then reduce to 5D/4D, we obtain massive 5D/4D representations with mass $M^2 = |Z|^2$. 
  
  Thus, Kaluza--Klein compactification of massless higher-dimensional superfield theories yields massive representations in lower dimensions that saturate the BPS bound $M_{4D} = |Z|$, known as short or BPS multiplets \cite{Moore}. Although this method produces only short multiplets with respect to $4D, \mathcal{N}=2$ supersymmetry, their subsequent decomposition into $4D, \mathcal{N}=1$ superfields yields generic massive $\mathcal{N}=1$ multiplets.
  
  Before proceeding to specific models, let us examine how the reduction operates on superspace coordinates and covariant derivatives. Since both $5D$ and $4D$ superspaces with central charges descend from $6D, \mathcal{N}=(1,0)$ harmonic superspace, it is convenient to start from $6D$ in a manifest $4+2$ split: the coordinate $x^5$ will serve as the Kaluza--Klein compactification direction, while $x^6$ generates the central charge required for the $5D, \mathcal{N}=1$ supergravity compensator in Section \ref{sec: 6}.

\medskip

\noindent\textbf{\underline{Central basis}}

\medskip

In the central basis 6D HSS is parametrized by by the following set of coordinates
$
	\mathbb{HR}^{6+2|8}  =	\left\{ x^{\alpha\dot{\alpha}}, x^5, x^6, \theta^i_\alpha, \bar{\theta}_{\dot{\alpha}i},  u^\pm_i \right\}
.$
Covariant spinor derivatives are given by ($x^{\alpha\dot{\alpha}} := \sigma_a^{\alpha\dot{\alpha}} x^a $, $\partial_{\alpha\dot{\alpha}} : = \frac{1}{2} \sigma^a_{\alpha\dot{\alpha}} \partial_a$):
\begin{equation}
	\begin{split}
		&\mathcal{D}^i_\alpha =
		+ \partial_\alpha^i 
		+
		2i \bar{\theta}^{\dot{\alpha}i} \partial_{\alpha\dot{\alpha}}
		-
		i \theta^i_\alpha ( \partial_5 - i \partial_6 ),
		\\
		& 
		\bar{\mathcal{D}}_{\dot{\alpha}i}
		=
		- \bar{\partial}_{\dot{\alpha}i} 
		-
		2 i 
		\theta^\alpha_i \partial_{\alpha\dot{\alpha}}
		-
		i\bar{\theta}_{\dot{\alpha}i} ( \partial_5 + i \partial_6 ).
	\end{split}	
\end{equation}	
Harmonic derivatives are  given  by
$\partial^{++},  \partial^{--},  \partial^0 . $
$5D$ analyticity condition in central basis is:
\begin{equation}\label{eq: 5D A CB}
	u^+_i \mathcal{D}^i_{\hat{\alpha}} \Phi_A(x, \theta, u) = 0.
\end{equation}	
To resolve this condition, it is convenient to choose the analytic basis where the $u^+$ projection of spinor derivative is short.

\medskip

\noindent\textbf{\underline{Analytic basis}}
\medskip

In the analytic basis, 6D HSS is parametrized by coordinates
$
	\mathbb{HR}^{6+2|8}  = 	\left\{ x_A^{\alpha\dot{\alpha}}, x_A^5, x_A^6, \theta^\pm_\alpha, \bar{\theta}_{\dot{\alpha}}^\pm,  u^\pm_i \right\},
$
where analytic coordinates are defined by
\begin{equation}\label{eq: analyt coord}
	\begin{split}
		&x_A ^{\alpha\dot{\alpha}} := x^{\alpha\dot{\alpha}} - 4i \theta^{\alpha(i} \bar{\theta}^{\dot{\alpha}j)} u^+_i u^-_j,
		\\
		&x^5_A := x^5 + i \left[ (\theta^{\alpha i}\theta^j_\alpha)
		-
		(\bar{\theta}^i_{\dot{\alpha}} \bar{\theta}^{\dot{\alpha}j}) \right] u^+_i u^-_j,
		\\
		&x^6_A := x^6 + \left[ (\theta^{\alpha i}\theta^j_\alpha)
		+
		(\bar{\theta}^i_{\dot{\alpha}} \bar{\theta}^{\dot{\alpha}j})  \right] u^+_i u^-_j,
		\\
		& \theta^{\pm}_{\hat{\alpha}} :=   \theta^{i}_{\hat{\alpha}} u^\pm_i.
	\end{split}
\end{equation}	
The analytic superspace is introduced as
$
	\mathbb{HA}^{6+2|4}  =	\left\{ x_A^{\alpha\dot{\alpha}}, x_A^5, x_A^6, \theta^+_\alpha, \bar{\theta}_{\dot{\alpha}}^+,  u^\pm_i \right\}.
$

\medskip

Covariant spinor derivatives in the analytic basis are given by  \cite{Ohta:1985ba}:  
\begin{equation}\label{eq: cov der}
	\begin{split}
		&\mathcal{D}^+_{\hat{\alpha}} = \partial^+_{\hat{\alpha}},
		\\
		&\mathcal{D}^-_\alpha = - \partial^-_\alpha + 4i \bar{\theta}^{-\dot{\alpha}}\partial_{\alpha\dot{\alpha}}
		-
		2i \theta^-_\alpha
		\left( \partial_5 - i \partial_6 \right),
		\\
		&\bar{\mathcal{D}}^-_{\dot{\alpha}} = - \partial^-_{\dot{\alpha}} - 4i \theta^{-\alpha}\partial_{\alpha\dot{\alpha}}
		-
		2i \bar{\theta}^-_{\dot{\alpha}} 	
		\left( \partial_5 + i \partial_6 \right).
	\end{split}
\end{equation}
In the analytic basis, $\mathcal{D}^+_{\hat{\alpha}}$ takes a short form. This allows us to solve the analyticity condition\eqref{eq: 5D A CB} as
\begin{equation}
	\mathcal{D}^+_{\hat{\alpha}} \Phi_A = 0 
	\qquad
	\Rightarrow
	\qquad
	\Phi_A =  \Phi_A (x_A, x_A^5, x_A^6,  \theta^+, u).
\end{equation}	 
Covariant harmonic derivatives in the analytic basis are long:
\begin{equation}\label{eq: 6D harmonic der}
	\begin{split}
		&\mathcal{D}^{++} = \partial^{++} - 4i \theta^{+\alpha} \bar{\theta}^{+\dot{\alpha}}\partial_{\alpha\dot{\alpha}}
		+
		\theta^{+\hat{\alpha}} \partial^+_{\hat{\alpha}}
+
		i\left[ (\theta^+)^2 - (\bar{\theta}^+)^2 \right] \partial_5
		+
		\left[ (\theta^+)^2 + (\bar{\theta}^+)^2 \right] \partial_6
		,
		\\
		&\mathcal{D}^{--} = \partial^{--} - 4i \theta^{-\alpha} \bar{\theta}^{-\dot{\alpha}}\partial_{\alpha\dot{\alpha}}
		+
		\theta^{-\hat{\alpha}} \partial^-_{\hat{\alpha}}
		+
		i\left[ (\theta^-)^2 - (\bar{\theta}^-)^2 \right] \partial_5
		+
		\left[ (\theta^-)^2 + (\bar{\theta}^-)^2 \right] \partial_6
		,
		\\
		& \mathcal{D}^0 = \partial^0.
	\end{split}	
\end{equation}

Upon reduction to five or four dimensions, the derivatives with respect to the extra coordinates are interpreted as real or complex central charges, respectively.

\section{Hypermultiplet reduction}\label{sec: 4}

Before turning to the compactification of gauge
theories, we briefly illustrate the reduction
procedure in the simpler case of the hypermultiplet.

We now compactify the physical coordinate $x^5$
on a circle of radius $r$.
A general analytic five-dimensional superfield
satisfying these conditions admits the following
Fourier expansion:
\begin{equation}
	\mathbb{q}^{+} (x^{\alpha\dot{\alpha}}_A, x_A^5, \theta^+, u)
	=
	\frac{1}{\sqrt{2\pi r}}
	\sum_{n\in \mathbb{Z}} e^{-i n\frac{x_A^5}{r}} q^{+}_n(x_A^{\alpha\dot{\alpha}}, \theta^+, u).	
\end{equation}	
The analyticity condition requires the $S^1$ harmonics to depend on the coordinate $x^5_A$. Then the modes $q^+_n$ in the expansion are analytic 4D superfields.

Substituting the expansion into the five-dimensional hypermultiplet action
$$
	S^{\text{hyper}}_{5D} [\mathbb{q}^+] 
	=
	\int d^5x d^4\theta^+ du \, 
	\tilde{\mathbb{q}}^+  \mathcal{D}^{++}_{5D}   \mathbb{q}^+
$$
 and integrating over the compact $x^5_A$ coordinate, we obtain
$
		S^{\text{hyper}}_{5D} [\mathbb{q}]
		=
	\sum_{n\in \mathbb{Z}} S^{\text{hyper}}_{4D} [q_n],
$
where 4D hypermultiplet action is given by
\begin{equation}
	S^{\text{hyper}}_{4D}   [q_n]	=
		\int d^4x d^4\theta du \,
		\tilde{q}^+_{n} \left( \mathcal{D}^{++} 
		+
		\left[ (\theta^+)^2 - (\bar{\theta}^+)^2 \right] \frac{n}{r}
		\right) q^+_n.
\end{equation}	
As expected, this yields the standard massive hypermultiplet action \cite{18, Ohta:1985ba}.
We see that $n=0$ mode corresponds to massless $4D$ hypermultiplet, whereas the
massive modes correspond to the tower of massive $4D$ hypermultiplets with $m_n^2 = \frac{n^2}{r^2}$.  All massive hypermultiplets are BPS, or short $\mathcal{N}=2$ supermultiplets, since by the construction $|m| = |Z|$.

\section{ $\mathcal{N}=2, D=4$ long massive vector multiplet}\label{sec: long}

In this section, we review the long massive
$4D,\mathcal{N}=2$ vector multiplet. This example
will serve as a point of comparison for the BPS
vector multiplet obtained through dimensional
reduction. It also illustrates how mass terms can
be incorporated into the harmonic-superspace
formalism. We focus on the superfield equations
of motion, while the component content and
component action are discussed in
\cite{Maslikov:1991yk, Maslikov:1992rn}.

\smallskip

Action of massive $4D, \mathcal{N}=2$ long vector multiplet is given by \cite{Galperin:1984av} \footnote{We note that this is the only possibility for massive term based on dimensional considerations
	$
	[V^{++}] = 0,
	[d^4x d^8\theta du] = 0,
	[d \zeta^{(-4)} ] = -2.$}:
\begin{equation}\label{eq: massive vector act}
	S^{\text{mass}}_{4D} [V^{++}] = \int d^4x d^8\theta du \; V^{++} V^{--} - 
	m^2 \int d \zeta^{(-4)} V^{++} V^{++}.
\end{equation}	
Here $V^{++}$ is the unconstrained  analytical prepotential, satisfying the reality condition under tilde-conjugation $\widetilde{V^{++}} = V^{++}$. The superfield $V^{--}$ is defined as the solution of harmonic zero-curvature equation:
\begin{equation}\label{eq: ZCE}
	\mathcal{D}^{++} V^{--} = \mathcal{D}^{--} V^{++}.
\end{equation}

The action \eqref{eq: massive vector act} can
also be rewritten in a gauge-invariant form
by introducing an analytic St\"uckelberg
superfield $\omega$. This is achieved by the
following field redefinition:
\begin{equation}
	V^{++} \to V^{++} - \mathcal{D}^{++} \omega.
\end{equation}	
which is invariant under spin-1 supergauge transformations  
$\delta V^{++} = \mathcal{D}^{++} \lambda,
\delta \omega = \lambda,
$
with $\lambda$ unconstrained analytic parameter. 

\medskip
Now we will show that this action indeed describes a massive vector supermultiplet.
The superfield equations of motion of the theory \eqref{eq: massive vector act} is given by
\begin{equation}\label{eq: SF eom spin 1}
	(\mathcal{D}^+)^4 V^{--} -  m^2 V^{++} = 0.
\end{equation}	
Then zero-curvature condition \eqref{eq: ZCE} 
implies the constraint on analytic prepotential 
$
\mathcal{D}^{++} V^{++} = 0.
$
As  a consequence we have for the solution of zero curvature equation  $V^{--} = \frac{1}{2} (\mathcal{D}^{--})^2 V^{++}$\footnote{This property is obvious in the central basis.}. After substitution of this solution  in eq. \eqref{eq: SF eom spin 1} we obtain\footnote{Here we used relation $(\mathcal{D}^+)^4 (\mathcal{D}^{--})^2 \Phi (\zeta) = -2 \Box  \Phi (\zeta)$ for arbitrary analytic superfield $ \Phi (\zeta)$, which is the direct consequence of covariant derivatives algebra.} massive Klein-Gordon equation:
$$
(\Box + m^2) V^{++} = 0.
$$

Equation  $\mathcal{D}^{++} V^{++} = 0$ leads to component expansion for the real analytic prepotential:
\begin{equation}
	\begin{split}
		V^{++} (\zeta) =\; &C^{ij} u^+_i u^+_j + \theta^{+\alpha} \chi^i_\alpha u^+_i + \bar{\theta}^{+\dot{\alpha}} \bar{\chi}^i_{\dot{\alpha}} u^+_i
		+
		i (\theta^+)^2 M - i (\bar{\theta}^+)^2 \bar{M}
		\\&-
		4i \theta^{+\alpha} \bar{\theta}^{+\dot{\alpha}} (A_{\alpha\dot{\alpha}} - \partial_{\alpha\dot{\alpha}} C^{(ij)}u^+_i u^-_j )
		\\&	+
		(\bar{\theta}^+)^2 \theta^{+\alpha} \lambda^i_\alpha u^-_i
		+
		(\theta^+)^2 \bar{\theta}^{+\dot{\alpha}} \bar{\lambda}^i_{\dot{\alpha}} u^-_i
		+ 
		(\theta^+)^4 m^2 C^{(ij)} u^-_i u^-_j.
	\end{split}
\end{equation}
Here all component fields  are  functions of $x_A$. Fermionic fields  satisfy equation  $	2i \partial_\alpha^{\;\dot{\alpha}} \bar{\chi}_{\dot{\alpha}}^i + \lambda^i_\alpha = 0$, vector field satisfy $\partial_{\alpha\dot{\alpha}} A^{\alpha\dot{\alpha}} = 0$. So we identify the set of physical fields:

\begin{itemize}
	\item Massive vector field $A_{\alpha\dot{\alpha}}$ with on-shell equations  $(\Box + m^2) A_{\alpha\dot{\alpha}} = 0,  \partial^{\alpha\dot{\alpha}}A_{\alpha\dot{\alpha}} = 0.$
	
	\item Two doublets Weyl spinor fields ($\psi^i_\alpha:= m\chi^i_\alpha, \lambda^i_\alpha $) with on-shell Dirac equations:
	\begin{equation*}
		\begin{cases}
			2i \partial_\alpha^{\;\dot{\alpha}} \bar{\psi}^i_{\dot{\alpha}} + m \lambda^i_\alpha = 0,
			\\		
			2i \partial_{\dot{\alpha}}^{\;\alpha} \lambda^i_\alpha - m \bar{\psi}^i_{\dot{\alpha}} = 0.
		\end{cases}
	\end{equation*}	
	So superfield equations paired two Weyl spinors ($\psi$ and $\lambda$) from different levels of the $\theta$-expansion into one massive Dirac fermion. The doubling occurred automatically due to the "swelling" of the superfield when the gauge freedom is removed.
	
	\item Five massive scalar fields: complex scalar field $M$ and real triplet $C^{(ij)}$ which satisfy KG equation.
	We note, that scalar triplet $D^{(ij)} $ is given by $D^{(ij)} = m^2 C^{(ij)}$.
\end{itemize}	
This is exactly on-shell content of long massive $4D, \mathcal{N}=2$ vector multiplet:
$$
\left\{ A_{\alpha\dot{\alpha}}, (\psi^i_\alpha, \lambda^i_\alpha), M, C^{ij}  \right\}.
$$
This multiplet contains $8_B + 8_F$ degrees of freedom.
Off-shell, this supermultiplet contains infinitely many d.o.f.

\medskip

\noindent \textbf{Massless limit}

\noindent It is highly instructive to discuss what happens in the massless limit. In the massless case using gauge transformations $\delta_\lambda V^{++} = \mathcal{D}^{++} \lambda$ one can fix gauge\footnote{This is possible because the operator $\left(\mathcal{D}^{++}\right)^2$ is invertible.} , which allows to rewrite massless equations of motions $(\mathcal{D}^+)^4 V^{--}$ is following form 
\begin{equation}
	\mathcal{D}^{++}V^{++} = 0
	\qquad
	\Rightarrow
	\qquad
	\begin{cases}
		\Box  V^{++} = 0,
		\\
		\mathcal{D}^{++}  V^{++} = 0. 
	\end{cases}	
\end{equation}
This gauge is a supersymmetric analogue of Lorenz gauge.
However, unlike the massive case
in this gauge there is the residual gauge freedom with analytic parameter $\lambda_{res}$ satisfying
$
\left(\mathcal{D}^{++} \right)^2 \lambda_{res} = 0.$

Using these residual gauge freedom we obtain  on-shell content of the massless $\mathcal{N}=2$ vector multiplet:
$$
\left\{A_{\alpha\dot{\alpha}}, \lambda^i_\alpha, M \right\},
\qquad 4_B+4_F.
$$

\section{Dimension reduction of $5D, \mathcal{N}=1$ Maxwell theory}\label{sec: 5}

We now apply the reduction procedure to the
$5D,\mathcal{N}=1$ vector multiplet. We work in
$5D$ harmonic superspace with a trivial central
charge and set $\partial_6\equiv 0$ throughout
this section.
We first perform the reduction of the vector
superfield and derive the corresponding
four-dimensional gauge structure. We then analyze
the equations of motion and identify the on-shell
component content. Finally, we reduce the action.

\smallskip

The harmonic-superspace action for the massless $5D, \mathcal{N}=1$ vector multiplet  is given by:
\begin{equation}\label{eq: 5D action}
	\begin{split}
	S^{\text{massless}}_{5D} 
&	=
	\int d^5x  d^8\theta du \, \mathbb{V}^{++}(x_A, \theta^+, u) \mathbb{V}^{--}(x_A, \theta^+, \theta^-, u).
	\end{split}
\end{equation}	
Here vector prepotential $\mathbb{V}^{++}(x_A, \theta^+, u)$ satisfies 5D analyticity condition \eqref{eq: 5D A CB}.
Superfield
$\mathbb{V}^{--}$ is the solution of the 5D zero-curvature equation 
\begin{equation}
	\mathcal{D}^{++}_{5D} \mathbb{V}^{--}(x, \theta, u)
	=
		\mathcal{D}^{--}_{5D} \mathbb{V}^{++}(x, \theta^+, u).
\end{equation}	
Here 5D harmonic derivatives are defined by \eqref{eq: 6D harmonic der} with $\partial_6 \to 0$. 
The action \eqref{eq: 5D action} is invariant under
gauge transformations of $5D$ vector prepotential $
	\delta_\lambda \mathbb{V}^{++} = \mathcal{D}_{5D}^{++} \bblambda$. 

The superfield equation of motion of the theory \eqref{eq: 5D action}  are given by
$
	(\mathcal{D}^+)^4 \mathbb{V}^{--} = 0.
$
These equations are cumbersome to analyze, as we need to solve the zero-curvature equations. A more convenient approach, as in $4D$ cases, is to pass to an equivalent system
\begin{equation}
		\begin{cases}
		\mathcal{D}^{++}_{5D} \mathbb{V}^{++}  = 0,
		\\
		\Box_{5D}  \mathbb{V}^{++}  = 0.
	\end{cases}	
\end{equation}	
This system of equations is defined up to the residual gauge freedom
$
	\delta \mathbb{V}^{++} = \mathcal{D}_{5D}^{++} \bblambda_{res}
$
 with parameter
$
 \left( 	\mathcal{D}_{5D}^{++} \right)^2 \bblambda_{res} = 0.
$

\medskip

We now compactify the fifth spacetime coordinate
\begin{equation}\label{eq: 5D fourier}
	\mathbb{V}^{++} (x^{\alpha\dot{\alpha}}_A, x_A^5, \theta^+, u)
	=
	\frac{1}{\sqrt{2\pi r}}
	\sum_{n\in \mathbb{Z}} e^{-i n\frac{x_A^5}{r}} \mathcal{V}^{++}_n(x_A^{\alpha\dot{\alpha}}, \theta^+, u).	
\end{equation}	
The resulting Fourier modes are complex $4D$ analytic
superfields, while the reality of the original
five-dimensional prepotential relates modes with
opposite Fourier numbers $\widetilde{V^{++}_n} = V^{++}_{-n}$.
Analogously, the expansion of the $5D$ analytic gauge parameter is given by
\begin{equation}
	\bblambda
	(x^{\alpha\dot{\alpha}}_A, x_A^5, \theta^+, u)
	=
	\frac{1}{\sqrt{2\pi r}}
	\sum_{n\in \mathbb{Z}} e^{- in\frac{x_A^5}{r}} \lambda_n(x_A^{\alpha\dot{\alpha}}, \theta^+, u).	
\end{equation}

\subsection{KK reduction and off-shell gauge structure}

For definiteness, we consider a positive Fourier mode
$n>0$ and denote its mass by $m=n/r$. In what follows,
we suppress the Fourier index on the corresponding
four-dimensional superfields. The induced gauge
transformation is
\begin{equation}\label{eq: BPS gauge transformations}
	\delta_\lambda\mathcal V^{++}
	=\mathcal D_m^{++} \callambda,
\end{equation}
where
$\mathcal D_m^{\pm\pm}
=\mathcal D^{\pm\pm}|_{\partial_5\to-im,\,
	\partial_6\to0}$.
Unlike the long massive multiplet considered in
Section~\ref{sec: long}, the mass enters through
the deformation of the harmonic derivatives rather
than through an explicit analytic mass term.

Using this gauge freedom one can impose following Wess-Zumino type gauge
\begin{equation}
	\begin{split}
	\mathcal{V}^{++} (\zeta) = 
	&(\theta^+)^2 M 
	-
	(\bar{\theta}^+)^2 \bar{N}
	-
	4i \theta^{+\alpha} \bar{\theta}^{+\dot{\alpha}} A_{\alpha\dot{\alpha}}
\\&	+
	(\theta^+)^2 \bar{\theta}^{+}_{\dot{\alpha}} \bar{\psi}^{\dot{\alpha}i}(x) u^-_i 
		+
	(\bar{\theta}^+)^2 \theta^{+\alpha} \chi_{\alpha}^{i}(x) u^-_i 
	+
	(\theta^+)^4 D^{(ij)}(x) u^-_i u^-_j
	\end{split}
\end{equation}	
with residual gauge freedom
\begin{equation}
	\delta A_{\alpha\dot{\alpha}}(x) = \partial_{\alpha\dot{\alpha}} a(x),
	\qquad
	\delta M (x) = m a(x),
	\quad
	\delta \bar{N}(x) = m a(x) .
\end{equation}	
The residual transformation identifies $M+\bar N$
as a Stückelberg combination, whereas the orthogonal
combination is gauge invariant. Since the nonzero
Fourier modes are complex, the component fields in
the Wess--Zumino gauge are complex as well. Taking
into account the residual complex gauge freedom,
the off-shell component content is:
\begin{equation}
	\left\{
	A_{\alpha\dot\alpha},
	\bar\psi_{\dot\alpha}^{\,i},
	\chi_\alpha^i,
	M,\bar N,D^{(ij)}
	\right\},
	\qquad
	16_B+16_F.
\end{equation}
In contrast, the long massive vector multiplet
contains an infinite set of off-shell harmonic
auxiliary fields.

\subsection{On-shell equations and component content}

Now we turn to the analysis of the physical degrees of freedom of the multiplet.
Equations
 on massless $5D$ vector multiplet imply for massive mode 
\begin{equation}\label{eq: mass mode eq}
	\begin{cases}
	\left( \Box + m^2 \right) 	\mathcal{V} ^{++} = 0,
	\\
	\mathcal{D}_m^{++} \mathcal{V}^{++} = 0,
	\end{cases}
\end{equation}
These equations are defined up to residual gauge transformations
$
	\delta_{\text{res}} 	\mathcal{V} ^{++} := \mathcal{D}_m^{++} 
	\lambda_{\text{res}} 
$
with parameter $\lambda_{res}$, which satisfies
$
	\left( \mathcal{D}_m^{++}  
	\right)^2
	\lambda_{\text{res}}  = 0.
$

To determine the on-shell component content, we solve
the equations \eqref{eq: mass mode eq} without fixing
the residual gauge freedom. The resulting general
analytic superfield contains both physical and
gauge-dependent components:
\begin{equation}\label{eq: on-shell expansion}
	\begin{split}
		\mathcal{V}^{++} (\zeta) = \,
		&
		C^{ij} u^+_i u^+_j
		+
		\theta^{+\alpha} \chi_\alpha^i u^+_i
		+
		\bar{\theta}^{+\dot{\alpha}} \bar{\rho}_{\dot{\alpha}}^i u^+_i
	\\&	+
		(\theta^+)^2 ( M - mC^{(ij)} u^+_i u^-_j )
		-
		(\bar{\theta}^+)^2 ( \bar{N} - mC^{(ij)} u^+_i u^-_j  )
	\\&	-
		4i \theta^{+\alpha} \bar{\theta}^{+\dot{\alpha}} \left(A_{\alpha\dot{\alpha}}
		- \partial_{\alpha\dot{\alpha}} C^{(ij)} u^+_i u^-_j
		\right)
		\\&	+
		(\theta^+)^2 \bar{\theta}^{+}_{\dot{\alpha}} \bar{\psi}^{\dot{\alpha}i} u^-_i 
		+
		(\bar{\theta}^+)^2 \theta^{+\alpha} \lambda_{\alpha}^{i} u^-_i 
		-
		 (\theta^+)^4 m^2 C^{(ij)} u^-_i u^-_j,
	\end{split}
\end{equation}	
where component fields satisfy Klein-Gordon equations and 
\begin{equation}\label{eq: fields eom}
	\begin{cases}
	\lambda_\alpha^i 
	+
	 2i \partial_\alpha^{\dot{\alpha}} \bar{\rho}^i_{\dot{\alpha}}
	+
	m \chi^i_\alpha = 0,
	\\
	\bar{\psi}^i_{\dot{\alpha}} + 2i \partial_{\dot{\alpha}}^{\alpha} \chi^i_\alpha 
	-
	m \bar{\rho}^i_{\dot{\alpha}} = 0,
	\\
	4 \partial^{\alpha\dot{\alpha}} A_{\alpha\dot{\alpha}} 
	- m \left( M + \bar{N} \right) = 0. 
\end{cases}
\end{equation}	
We see that first and second equations imply that Weyl fields $\lambda$ and $\psi$ satisfy Dirac equation
\begin{equation}\label{eq: D eq BPS}
\begin{cases}	
	2i \partial_{\dot{\alpha}}^\alpha \lambda_\alpha^i - m  	\bar{\psi}_{\dot{\alpha}}^i = 0 ,
	\\
	 2i \partial_\alpha^{\dot{\alpha}} \bar{\psi}^i_{\dot{\alpha}}
	+
	m	\lambda_\alpha^i 
	=
	0.
\end{cases}	
\end{equation}	

The residual gauge transformations act on the scalar
fields as
\begin{equation}
	\delta(M+\bar N)=2ma,
	\qquad
	\delta(M-\bar N)=0.
\end{equation}
Thus, $M+\bar N$ is the Stückelberg field and can
be eliminated by a gauge choice. The third equation
in \eqref{eq: fields eom} then imposes the
transversality condition
$\partial^{\alpha\dot\alpha}A_{\alpha\dot\alpha}=0$.
Consequently, $A_{\alpha\dot\alpha}$ describes a
massive complex vector, while the gauge-invariant
combination
$
	\phi=M-\bar N
$
is a physical complex scalar.
Similarly we gauge out part of fermionic d.o.f.
The remaining fermionic degrees of freedom are
described by one massive Dirac doublet
$(\lambda_\alpha^i,\bar\psi_{\dot\alpha}^i)$.
We therefore obtain the on-shell field content
\begin{equation}
	\left\{
	A_{\alpha\dot\alpha},
	\lambda_\alpha^i,
	\bar\psi_{\dot\alpha}^i,
	\phi
	\right\},
	\qquad 8_B+8_F.
\end{equation}
It is instructive to compare this spectrum with that of the long massive vector multiplet in Section~\ref{sec: long}. The long multiplet is described by a \textit{real} analytic prepotential $V^{++}$ and carries $8_B + 8_F$ real on-shell degrees of freedom (one real massive vector, five real scalars, and two massive Weyl spinor doublets). By contrast, a massive BPS multiplet carries a non-vanishing central charge $Z_n = m_n$ and is therefore inherently \textit{complex} (combining modes $+n$ and $-n$). Its $8_B + 8_F$ real degrees of freedom correspond to $4_B + 4_F$ complex states: one complex massive vector ($6_B$), one complex scalar $\phi$ ($2_B$), and one massive Dirac doublet ($8_F$), which is precisely half the field content of a complexified long multiplet.

\subsection{Reduced action}

To reduce the $5D$ action \eqref{eq: 5D action}, one need to solve zero-curvature equation. We represent the solution as following Fourier series 
\begin{equation}
	\mathbb{V}^{--}  (x_A^{\alpha\dot{\alpha}}, x^5_A, \theta^\pm, u)
	=
		\frac{1}{\sqrt{2\pi r}}
	 \sum_{n\in \mathbb{Z}}
	e^{-i n\frac{x_A^5}{r}} \mathcal{V}^{--}_n(x_A^{\alpha\dot{\alpha}}, \theta^\pm , u).
\end{equation}	
Here  $4D$ modes are defined by massive zero-curvature condition equation 
$
\mathcal{D}_{m_n}^{++}
	V^{--}_n
	=
\mathcal{D}_{m_n}^{--}
 V^{++}_n.
$
Then we obtain for reduced action
\begin{equation}
	\begin{split}
	S^{\text{massless}}_{5D}[\mathbb{V}^{++}] =
 \int d^4x d^8\theta du \, 
 \left( \mathcal{V}^{++}_0  \mathcal{V}^{--}_{0} 
	+
	2  \sum_{n=1}^\infty   \mathcal{V}^{++}_n  \tilde{\mathcal{V}}^{--}_{n} 
	\right).
	\end{split}
\end{equation}

The zero mode reproduces the standard massless
four-dimensional vector multiplet action. For each
positive Fourier mode the action has the same form, with difference that superfields are complex  and are related by the massive
harmonic zero-curvature equation. The action is
invariant under the gauge transformations
\eqref{eq: BPS gauge transformations}.

An equivalent representation of the action for BPS modes  is
\begin{equation}
	\begin{split}
		S_{4D}^{\rm mass}
		={}&
		\int d^4x\,d^8\theta\,du_1du_2\,
		\exp\left\{
		-im\left[x_A^5(u_1)-x_A^5(u_2)\right]
		\right\}
		\times
		\frac{
			\mathcal V^{++}(x,\theta,u_1)
			\widetilde{\mathcal V}^{++}(x,\theta,u_2)
		}{
			(u_1^+u_2^+)^2
		}.
	\end{split}
\end{equation}
This form makes explicit the dependence on the
difference of the analytic coordinates associated
with two harmonic points.

\section{Poincar\'e supergravity reduction}\label{sec: 6}

We now apply the same reduction procedure to
linearized $5D,\mathcal{N}=1$ Poincare supergravity.
In the harmonic-superspace formulation, this theory
is described in terms of a covariant harmonic
derivative of the following form\footnote{The vector compensator $h^{++6}$ is a vielbein associated with the central charge $\partial_6$, obtained by reduction from $6D$ theory. It would be interesting to construct a linearized six-dimensional supergravity theory and study dimensional reduction to $5D$ . We expect its structure to differ from that of the actions in $4D$ and $5D$, since tensor compensators are required for the formulation \cite{Sokatchev:1988aa}.}:
\begin{equation}
	\begin{split}
	\mathfrak{D}_{5D}^{++} :=
	\mathcal{D}_{5D}^{++} + \hat{\mathbb{G}}^{++}
	&=
		\mathcal{D}_{5D}^{++}
	+
	\mathbb{h}^{++\hat{a}} \partial_{\hat{a}}
	+
	\mathbb{h}^{++\hat{\alpha}+} \partial^-_{\hat{\alpha}}
	+
	\mathbb{h}^{++6}  \partial_6,
	\\&
	=
	\mathcal{D}_{5D}^{++}
	+
	\mathbb{G}^{++\hat{a}} \partial_{\hat{a}}
	+
		\mathbb{G}^{++\hat{\alpha}+} \mathcal{D}^-_{\hat{\alpha}}
	+
		\mathbb{G}^{++6}  \partial_6.
	\end{split}
\end{equation}	
Here $\hat a=(a,5)$ denotes the five-dimensional
spacetime index, while $x^6$ is the central-charge
coordinate. The supergravity prepotentials and gauge
parameters are independent of the central-charge
coordinate. The prepotential $\mathbb h^{++6}$ is nevertheless
retained as the vielbein associated with the
central-charge generator $\partial_6$. The covariant harmonic derivative satisfies analyticity condition
$
	[\mathcal{D}^+_{\hat{\alpha}},	\mathfrak{D}_{5D}^{++}  ]
	=
	0.
$

The covariant harmonic derivative admits two
equivalent representations. The first is expressed
in terms of the analytic prepotentials $\mathbb h$,
while the second involves the supersymmetry-invariant
superfields $\mathbb G$. The latter are generally
non-analytic and can be straightforwardly constructed from the
unconstrained analytic prepotentials.

\smallskip

Gauge supertransformations  are given by
\begin{equation}
	\delta_{\lambda}\hat{\mathbb{G}}^{++} = [\mathcal{D}_{5D}^{++}, \hat{\mathbb{\Lambda}}],
\end{equation}	
where the analytic first-order differential operator is given by
\begin{equation}
	\hat{\mathbb{\Lambda}} = \bblambda^{\hat{a}} \partial_{\hat{a}}  
	+
	\bblambda^{+\hat{\alpha}} \partial^-_{\hat{\alpha}} 
	+
	\bblambda^6 \partial_6
	=
	\mathbb{\Lambda}^{\hat{a}} \partial_{\hat{a}} 
	+
	\mathbb{\Lambda}^{+\hat{\alpha}} \mathcal{D}^-_{\hat{\alpha}} 
	+
	\mathbb{\Lambda}^6 \partial_6.
\end{equation}	
Where $\bblambda$ parameters are analytic superfields.
Component counterpart of this operator transformation law in given by
\begin{equation}\label{eq: gauge tr sg}
	\begin{split}
		\delta_\lambda \mathbb{G}^{++\hat{a}} = \mathcal{D}_{5D}^{++} \mathbb{\Lambda}^{\hat{a}},
		\qquad
			\delta_\lambda \mathbb{G}^{++\hat{\alpha}+} = \mathcal{D}_{5D}^{++} \mathbb{\Lambda}^{+\hat{\alpha}},
			\qquad
				\delta_\lambda \mathbb{G}^{++6} = \mathcal{D}_{5D}^{++} \mathbb{\Lambda}^{6}.
	\end{split}	
\end{equation}	

\medskip

According to Ref. \cite{Buchbinder:2025yef}, gauge-invariant action for $5D, \mathcal{N}=1$ linearized Poincar\'e supergravity is given by (in metric signature $(+,-,-,-,-)$):
\begin{equation}
	S_{5D} = - \int d^5x d^8\theta du \left( \mathbb{G}^{++\hat{a}} \mathbb{G}^{--}_{\hat{a}} + 3 \mathbb{G}^{++6} \mathbb{G}^{--6}  \right). 
\end{equation}	
The negative-charge superfields are determined by
the harmonic zero-curvature condition $\mathcal{D}^{++}_{5D} \mathbb{G}^{--\hat{a},6} = \mathcal{D}^{--}_{5D} \mathbb{G}^{++\hat{a},6} $.

We now compactify the physical coordinate $x^5$
on a circle of radius $r$. All supergravity
prepotentials, gauge parameters, and
supersymmetry-invariant superfields are expanded
into Fourier modes, following the prescription
given in Eq. \eqref{eq: 5D fourier}.
The reality conditions of the parent
five-dimensional theory relate modes with
opposite Fourier numbers.

For a fixed Fourier mode, the derivative along the
compactified direction acts as
$
	\partial_5\longrightarrow -im_n,$
with
$
	m_n=\frac{n}{r}.
$
We denote the resulting massive harmonic derivatives
by
\begin{equation}
	\mathcal{D}_{m_n}^{\pm\pm}
	:=
	\left.
	\mathcal D_{5D}^{\pm\pm}
	\right|_{\partial_5\to-im_n}.
\end{equation}
The central-charge direction $\partial_6$ is retained
in the supergravity covariant derivatives, with the
prepotentials being independent of $x^6$.

Reduction procedure gives
\begin{equation}
	\begin{split}
	\mathbb{h}^{++\hat{a}} 
	\quad
	\to
	\quad
	(\mathbb{h}^{++a}, \mathbb{h}^{++5}),
\qquad
	\mathbb{h}^{++\hat{\alpha}+} 
	\;
	\to
	\quad
	(\mathbb{h}^{++\alpha+}, - \bar{\mathbb{h}}^{+++}_{\dot{\alpha}})
	\end{split}
\end{equation}	
We also decompose all analytic prepotentials and superparameters into analytic modes on compact $S^1$, which corresponds to $x_A^5$ coordinate.
After substituting the Fourier expansions into the
five-dimensional action and integrating over the
compact coordinate, we obtain
$
	S_{5D}=S_0^{SG} + 2\sum_{n=1}^\infty S_n^{\rm SG},
$
where
\begin{equation}\label{eq:sg_massive_action}
	\begin{split}
		S_n^{\rm SG}=-\int d^4x\,d^8\theta\,du\,
		\Big(
		&\mathcal G_n^{++a}
		\widetilde{\mathcal G}_{n,a}^{--}
		-\mathcal G_n^{++5}
		\widetilde{\mathcal G}_{n}^{--5}
		+3\mathcal G_n^{++6}
		\widetilde{\mathcal G}_{n}^{--6}
		\Big).
	\end{split}
\end{equation}
The negative-charge superfields are determined
by the massive zero-curvature condition. This action is invariant under gauge transformations 
$
	\delta \mathcal{G}^{\pm\pm a, 5, 6, \hat{\alpha}+ }_n = \mathcal{D}^{\pm\pm}_{m_n} \Lambda^{a, 5, 6, \hat{\alpha}+}.
$

For the zero mode, $m_0=0$ and the reduced action
describes a massless $4D$, $\mathcal{N}=2$ supergravity
sector. The first term contains the four-dimensional
gravitational prepotential, while the second term
describes an additional vector multiplet $\mathcal{G}^{++5}_0$ (supersymmetric Kaluza-Klein graviphoton) originating
from the fifth component of the five-dimensional
vielbein. The last term corresponds to the
$\mathcal{N}=2$ vector compensator associated with
the central-charge direction. Its relative coefficient
and sign are inherited from the five-dimensional
action.
Thus, the massless sector obtained by dimensional
reduction contains the standard supergravity and
compensator system together with an additional
Kaluza--Klein vector multiplet. In this respect, it
differs from the formulation of Ref.~\cite{Zupnik:1998td, Buchbinder:2021ite}
by the presence of this additional vector multiplet.

For the non-zero Fourier modes ($n \ge 1$), the action \eqref{eq:sg_massive_action} describes massive $4D, \mathcal{N}=2$ BPS supergravity multiplets with masses $m_n = \frac{n}{r} = |Z_n|$. The superfields $\mathcal{G}_n^{++5}$ and $\mathcal{G}_n^{++6}$ provide the St\"uckelberg degrees of freedom required to render the $4D, \mathcal{N}=2$ supergravity multiplet massive.

\section{Conclusion}\label{sec: 7}

In this paper, we have considered a Kaluza--Klein
reduction procedure in harmonic superspace and
applied it to the construction of massive
$4D,\mathcal{N}=2$ BPS supermultiplets.

First, we reduced the free $5D,\mathcal{N}=1$
Maxwell theory and obtained an off-shell superfield
description of the massive BPS vector multiplet.
To our knowledge, such a formulation has not
previously been presented in harmonic superspace.
We analyzed its off-shell gauge structure and
on-shell component content and compared it with
the long massive $4D,\mathcal{N}=2$ vector
multiplet. We then applied the same reduction
procedure to linearized $5D,\mathcal{N}=1$
Poincare supergravity and derived the corresponding
superfield action for the massive BPS
$4D,\mathcal{N}=2$ supergravity multiplet.

The technical simplicity of the approach stems
from the structure of gauge theories in harmonic
superspace, where gauge transformations act on
analytic prepotentials through harmonic
derivatives. This structure allows the
Kaluza--Klein reduction to preserve gauge
invariance, manifest supersymmetry, and
Grassmann analyticity, leading to gauge-invariant
off-shell formulations of short massive
supermultiplets.

The results obtained suggest several possible
directions for further investigation. We briefly
discuss two of them.

\smallskip

\textbf{1. }\textit{Generalization to higher-spin theories}

\smallskip

The proposed procedure can be extended to the
dimensional reduction of $5D,\mathcal{N}=1$
higher-spin theories \cite{Buchbinder:2025yef},
with the aim of constructing massive
$4D,\mathcal{N}=2$ BPS higher-spin supermultiplets.
The extension
to higher-spin theories does not require new
conceptual ingredients, although the increasing
complexity of the superfield and gauge structures
makes the corresponding calculations
combinatorially involved.
The oscillator formalism may prove particularly
useful in this context
\cite{Pashnev:1989gm, Gates:2005su}.
As a next step, it is interesting to reformulate
the resulting theories in $4D,\mathcal{N}=1$
superspace. This will provide a route to
gauge-invariant descriptions of massive
$\mathcal{N}=1$ higher-spin theories.

\smallskip

\textbf{2. }\textit{Long massive $\mathcal{N}=2$
	supergravity and higher-spin multiplets}

\smallskip

By construction, the Kaluza--Klein reduction
considered in this paper produces only BPS
supermultiplets and therefore cannot directly
describe long massive $\mathcal{N}=2$ multiplets.
As illustrated by the vector multiplet example,
the off-shell formulation of long massive
multiplets is expected to have a more involved
structure.

In particular, the general superfield equations
that select irreducible representations of the
$4D,\mathcal{N}=2$ super-Poincare algebra remain
to be constructed. This contrasts with the
$\mathcal{N}=1$ case, where such equations have
been known for a long time
\cite{Gates:1983nr, Gates:2013rka}.
We expect that further investigation of the
harmonic structure of the Casimir operators of
the $\mathcal{N}=2$ super-Poincare algebra,
initiated in \cite{Eremeev:2026yqj}, will provide
useful insights into this problem.
Ultimately, this may lead to the construction of
corresponding superfield equations and
gauge-invariant formulations of long massive
supermultiplets.

\acknowledgments

The author is grateful to E.~Ivanov for many fruitful discussions and his support. 
I am also pleased to thank D. Galakhov for discussion of BPS supermultiplets, A. Marshakov for discussion of massive $\mathcal{N}=2$ theories, and P. Petriakova for her interest in this work.
This work was partially supported by
the Foundation for the Advancement of Theoretical Physics and
Mathematics ``BASIS'', grant \verb|#| 25-1-1-10-4 .

\end{document}